\documentclass[12pt]{article} 
\usepackage{graphicx}
\usepackage{epsfig}
\usepackage{amsbsy}

\begin{document}
\baselineskip 10mm

\centerline{\large \bf Stability of C$_{20}$ fullerene chains}

\vskip 6mm

\centerline{L. A. Openov$^{*}$, I. V. Davydov, and A. I. Podlivaev}

\vskip 4mm

\centerline{\it Moscow Engineering Physics Institute (State
University), 115409 Moscow, Russia}

\vskip 2mm

$^{*}$ E-mail: LAOpenov@mephi.ru

\vskip 8mm

\centerline{\bf ABSTRACT}

The stability of (C$_{20}$)$_N$ chains, where C$_{20}$ fullerenes are covalently coupled is analyzed by numerical
simulation using a tight-binding potential. Various channels of losing the chain–cluster structure of the
(C$_{20}$)$_N$ complexes have been determined including the decay of the C$_{20}$ clusters, their coalescence,
and the separation of one C$_{20}$ fullerene from a chain. The lifetimes of the (C$_{20}$)$_N$ chains with
$N=$ 3 - 7 for $T=$ (2000 - 3500) K are directly calculated by the molecular dynamics method. It has been shown
that although the stability of the chains decreases with an increase in $N$, it remains sufficiently high even for
$N\gg 1$. An interesting lateral result is the observation of new (C$_{20}$)$_N$ isomers with the combination of
various intercluster bonds and the maximum binding energy of fullerenes in the chain.

\newpage

Since the experimental detection of the smallest possible
fullerene (C$_{20}$)$_N$ [1] in 2000 (see Fig. 1a), a problem of
existing a solid phase (fullerite) based on this fullerene
attracts considerable interest (by analogue with the known
fullerite consisting of C$_{60}$ fullerenes [2,3]). Theoretical
calculations [4-7] indicate the possibility of forming a condensed
state of C$_{20}$ fullerenes, although there is yet no commonly
accepted opinion about the crystal structure of this hypothetical
cluster substance [5-7]. In experiments, in addition to the
C$_{20}$ fullerene, only charged dimers (C$_{20}$)$_2^+$, as well
as complexes (C$_{20}$)$_N^+$ with $N=$ 3 - 13, were observed [8].
Reports [9,10] on synthesizing C$_{20}$ crystals have not yet been
corroborated. In contrast to the C$_{60}$ fullerite, where single
clusters are bounded due to weak van der Waals attraction, the
intercluster bonds in the C$_{20}$ fullerite must be covalent
according to the theory developed in [4-7]; therefore, they must
be very rigid. On the one hand, this can promote the stability of
the solid phase. On the other hand, the strong coupling between
clusters can lead to transformation into another atomic
configuration,

Previously, we showed [11] that a single C$_{20}$ fullerene loses
its spheroidal shape at high temperatures through decay into
various energetically unfavorable quasi-one-dimensional or
quasi-two-dimensional configurations rather than through a
transition to a "bowl" isomer with minimum total energy $E$. The
(C$_{20}$)$_2$ dimer can lose its molecular structure (see Fig.1b)
due both to the decay of one of the C$_{20}$ fullerenes and to the
coalescence of two C$_{20}$ fullerenes into a C$_{40}$ cluster
[12,13]. In this case, the height $U$ of the lowest energy barrier
bounding the initial metastable configuration decreases from
$\approx 5$ eV to $\approx 2.5$ eV [12]. This decrease leads to a
decrease in the activation energies of the corresponding processes
and, as a consequence, a decrease in the lifetime at a given
temperature [13]. With increasing the number of fullerenes in the
(C$_{20}$)$_N$ complexes with $N>$ 2, the stabilization of the
metastable state due to an increase in the number of nearest
neighbors, as well as appearance of new mechanisms of stability
loss, can be expected.

In this work, we theoretically analyze the thermal stability of
the quasi-one-dimensional (C$_{20}$)$_N$ chains as a first step of
analysis of the stability of macroscopic three-dimensional cluster
structures based on C$_{20}$ fullerenes. Such metastable chains
with various bonds between fullerenes (see Figs. 2a-c) were
previously considered using the density functional theory (DFT)
and the combination of the tight-binding approximation with the
DFT (DFTB method) [4,7]. It was shown that the binding energy of
fullerenes $\Delta E=E$(C$_{20}$) - $E[($C$_{20}$)$_N]/N$ for a
given $N$ is minimal (the total energy $E[($C$_{20}$)$_N]$ is
maximal) in chains with {\it open}-[2+2] intercluster bonds, as in
the isomer (C$_{20}$)$_2$ with the minimum total energy
$E[($C$_{20}$)$_2]$ (see Fig. 1b).

We use the tight-binding method [14], which was successfully
applied to simulate both small clusters and macroscopic carbon
structures (previously, we successfully used it to investigate
various characteristics of the C$_{20}$ and C$_{60}$  fullerenes,
as well as other carbon systems [11-13,15,16]]. First, we verify
that this method gives the same sequence of the (C$_{20}$)$_N$
chains in their energies (i.e., in $\Delta E$) as that obtained in
more accurate DFT calculations [4,7]. In agreement with [4,7], we
found that $\Delta E$ is maximal in the chains with {\it
open}-[2+2] bonds and is smaller in the chains with "twisted"
bonds (in terminology accepted in [7]). The extrapolation of data
obtained for $N=$ 3 - 12 to $N\rightarrow\infty$ gives $\Delta E=$
4.341, 4.338, and 2.744 eV/C$_{20}$ for {\it open}-[2+2], twisted,
and [2+2] bonds, respectively, see Fig. 3.

Moreover, we found that combinations of various intercluster bonds are possible, see Figs. 2d and 2e. In this case,
the binding energies $\Delta E$ are usually between the corresponding $\Delta E$ values for the chains with one
type of bonds (see Fig. 2d). However, there is an interesting exception. For $N\ge 4$, the presence of one or several
twisted bonds in the chain with {\it open}-[2+2] bonds can increase $\Delta E$, i.e., reduce the total energy. Thus, the
combination of {\it open}-[2+2] and twisted bonds are most energetically favorable. However, the energy gain is so
small (cf. Figs. 2a and 2e) that the configurations with pure {\it open}-[2+2] bonds and with the combination of
{\it open}-[2+2] and twisted bonds are almost degenerate.

To analyze the thermal stability of the (C$_{20}$)$_N$ chains, we
use the molecular dynamics method. Configurations with {\it
open}-[2+2] bonds are taken as initial configurations. At the
initial time, each atom is given a random velocity and
displacement so that the total momentum and angular momentum of
the entire system are equal to zero. Then, forces acting on atoms
are calculated and classical Newtonian equations of motion are
solved with a time step of $t_0=2.72\cdot 10^{-16}$ s. The total
energy of the system (the sum of the potential and kinetic
energies) remains unchanged in the simulation process, which
corresponds to the case, where the system is not in the thermal
equilibrium with the environment [15]. In this case, the dynamic
temperature $T$ is a measure of the relative motion energy of
atoms and is calculated as [17,18] $\langle
E_{kin}\rangle=\frac{1}{2}k_B T(3n-6)$, where $\langle
E_{kin}\rangle$ is the time-average kinetic energy of the entire
system, $k_B$ is the Boltzmann constant, and $n = 20N$ is the
number of atoms in a (C$_{20}$)$_N$ chain. Since the lifetimes
$\tau$ of the (C$_{20}$)$_N$ complexes to the time of loss of
their cluster-chain structure increase exponentially as $T$
decreases and, moreover, the time necessary for numerical
calculations for a given $T$  value increases with $N$ as $N^3$,
we perform simulation for the chains with $N\leq 7$ for
temperatures $T=$ (2000-3500) K. Nevertheless, we trace the
evolution of these chains for $t>$ 1 ns (and even for $t>$ 10 ns
for $N=3$), whereas {\it ab initio} calculations are usually
restricted to $t<$ 0.1 ns.

First, in the process of evolution of the  (C$_{20}$)$_N$ chains
for $t<\tau$, we observed very frequent (in time 0.1-1 ps) changes
of one or several {\it open}-[2+2] bonds between the C$_{20}$
fullerenes to twisted bonds and vise versa. These changes occur
due, first, to the closeness of the energies of the corresponding
configurations (see Fig. 2) and, second, to a low height of the
potential barrier separating these configurations ($U\approx 0.6$
eV for $N=2$ [12]), which is easily overcome at temperatures $T>$
2000 K. Such changes in the intercluster bonds lead neither to the
loss of the spheroidal shape of the C$_{20}$ fullerenes in the
chain nor to the breakdown of the chain.

We surprisingly observed one more way for "reorganizing" bonds
between the C$_{20}$ clusters, in which the overall chain shape of
the (C$_{20}$)$_N$ complexes also remains unchanged. In this case,
one of the {\it open}-[2+2] bonds is first broken and, then, is
recovered in a very short time ($\sim 0.1$ ps), but between other
atoms, being rotated at an angle of about 70$^0$ with respect to
the other {\it open}-[2+2] bonds in the chain, see Fig. 4a. Such
changes in the intercluster bonds occur quite rarely (likely
because the corresponding energy barrier is comparatively high);
nevertheless, we repeatedly observe them in the chains with
various $N$ values. It is interesting that the binding energies
$\Delta E$ of the (C$_{20}$)$_N$ isomers appearing in this case
are slightly higher than the corresponding $\Delta E$ values in
the isomers with {\it open}-[2+2] bonds or combinations of {\it
open}-[2+2] and twisted bonds (e.g., $\Delta E$ increases by 0.017
and 0.015 eV/C$_{20}$ for $N=3$ and 5, respectively, if the {\it
open}-[2+2] bond between the first and second C$_{20}$ fullerenes
in the chain is turned). We also observe the formation of isomers
in which one of the {\it open}-[2+2] bonds is not only turned, but
also twisted, see Fig. 4b. Their binding energies are as a rule
higher by a very small value of $\sim 0.01$ eV/C$_{20}$.
Therefore, these isomers correspond to energetically favorable
configurations for the corresponding $N$ values. We do not know
works reported about such (C$_{20}$)$_N$ isomers.

The problem of the optimal (for the minimum total energy) number
of turned and twisted bonds and the order of their alternation
with each other and with the {\it open}-[2+2] bonds requires a
separate consideration, which is beyond the scope of this paper.
This problem is insignificant for the main aim of this work, i.e.,
analysis of the thermal stability of (C$_{20}$)$_N$ chains,
because all found isomers are almost degenerate and hold the chain
structure. Nevertheless, note that the results obtained in this
work illustrate how the simulation of the dynamics of a complex
many-particle system allows one to "automatically" find new
metastable configurations the existence of which is not a priori
obvious and which can be missed when they are sought on the basis
of other (e.g., symmetry) considerations. {\it Ab initio}
calculations of the binding energies of the new (C$_{20}$)$_N$
isomers found in this work would be of interest.

Similar to the simulation of the thermal stability of the
(C$_{20}$)$_2$ cluster dimers [12,13], we observe the decays of
one of the fullerenes in the chain, see Fig. 5. In this case, the
binding energy $\Delta E$ decreases by 1.5 - 2.5 eV; i.e., the
system transits to an energetically less favorable configuration;
this is accompanied by a decrease in its temperature, because the
total energy is a constant. When the end fullerene decays (see
Fig. 5a), the fullerenes remaining in the chain hold their
individualities for a longer time than in the case of the decay of
one of the central fullerenes (Fig. 5b). This is explained
partially by a more significant decrease in $\Delta E$ (i.e., by a
stronger cooling of the system) and partially by a smaller number
of stressed covalent bonds in the remaining fullerenes. In any
case, soon after the decay of one fullerene, the (C$_{20}$)$_N$
complex completely loses its chain structure and transits to
various (usually, quasi-two-dimensional) configurations with lower
binding energies.

Similar to the (C$_{20}$)$_2$ dimer [12,13], in addition to decay,
there is a fundamentally different mechanism of losing the
individuality of the C$_{20}$ fullerenes in the chain: coalescence
of two neighboring C$_{20}$ fullerenes into a C$_{40}$ cluster,
see Fig. 6. In this case, the binding energy $\Delta E$ increases
by 0.5 - 1.5 eV (the larger increase corresponds to the C$_{40}$
cluster whose structure is closer to the ideal C$_{40}$
fullerene). An increase in $\Delta E$ is accompanied by an
increase in the temperature of the system, which leads to a fast
decay of both the C$_{40}$ cluster and the C$_{20}$ fullerenes
remaining in the chain. As a result, some quasi-two-dimensional
configuration is usually formed.

The separation of one C$_{20}$  fullerene from the (C$_{20}$)$_N$
chain (see Fig. 7) is a new channel of the breakdown of this chain
that is absent for the (C$_{20}$)$_2$ dimer. We emphasize that
such separation was observed only once for the chain with $N=3$.
This likely is explained by a small frequency factor $A$ of the
corresponding process. It is interesting that the separation of
the C$_{20}$ fullerene from the chain occurs after its transition
to an isomer with the turned {\it open}-[2+2] bond, cf. Figs. 4a
and 7.

In order to determine the activation energies $E_a$ for decay and
coalescence by the Arrhenius formula for $\tau(T)$ from the
numerical experiment, it is necessary to collect a sufficiently
large statistical sample for each $N$ value, as was previously
done for the C$_{60}$ fullerene [15] and (C$_{20}$)$_2$ dimers
[13]. Since the corresponding calculations require very large
computation time, we compare $\tau$ values obtained for certain
temperatures in the range $T=$ (2000 - 3500) K to the dependences
$\tau (T)$ found previously [13] for the decay and coalescence of
the C$_{20}$ fullerenes in the (C$_{20}$)$_2$ dimer, see Fig. 8.
In this case, similar to [13], the finite-heat-bath correction
[19,20] is taken into account by setting
$\tau^{-1}(T)=A\exp(-E_a/k_BT^*)$, where $T^*=T-E_a/2C$ [the
microcanonical heat capacity is assumed to be $C=(3n-6)k_B$, where
$n=20N$]. As seen in Fig. 8, the loss of the stability of the
chains for $T<2500$ K is caused only by the coalescence of the
C$_{20}$ fullerenes, whereas this loss for $T>2500$ K is caused by
both coalescence and decay, although the decay events occur much
rarely. Although an increase in $N$ leads to a decrease in $\tau$,
the (C$_{20}$)$_N$ chains remain sufficiently stable for $N\gg 1$.
The average coalescence activation energy for the (C$_{20}$)$_2$
dimer is equal to $E_a=$ 2.7 eV [13], this energy for the
(C$_{20}$)$_N$ chains with $N>$ 2 is only halved according to
analysis of the "computer experiment" data. Then, a
macroscopically large lifetime $\tau \sim 10^7$ s is obtained for
temperature $T=300$ K with the parameter $A \sim 10^{16}$ s$^{-1}$
[13].

Our next aim is to analyze the stability of two-dimensional and
three-dimensional systems based on the C$_{20}$ fullerenes, which
will be done in a future work.

\newpage

\vskip 20mm

\includegraphics[width=\hsize,height=17cm]{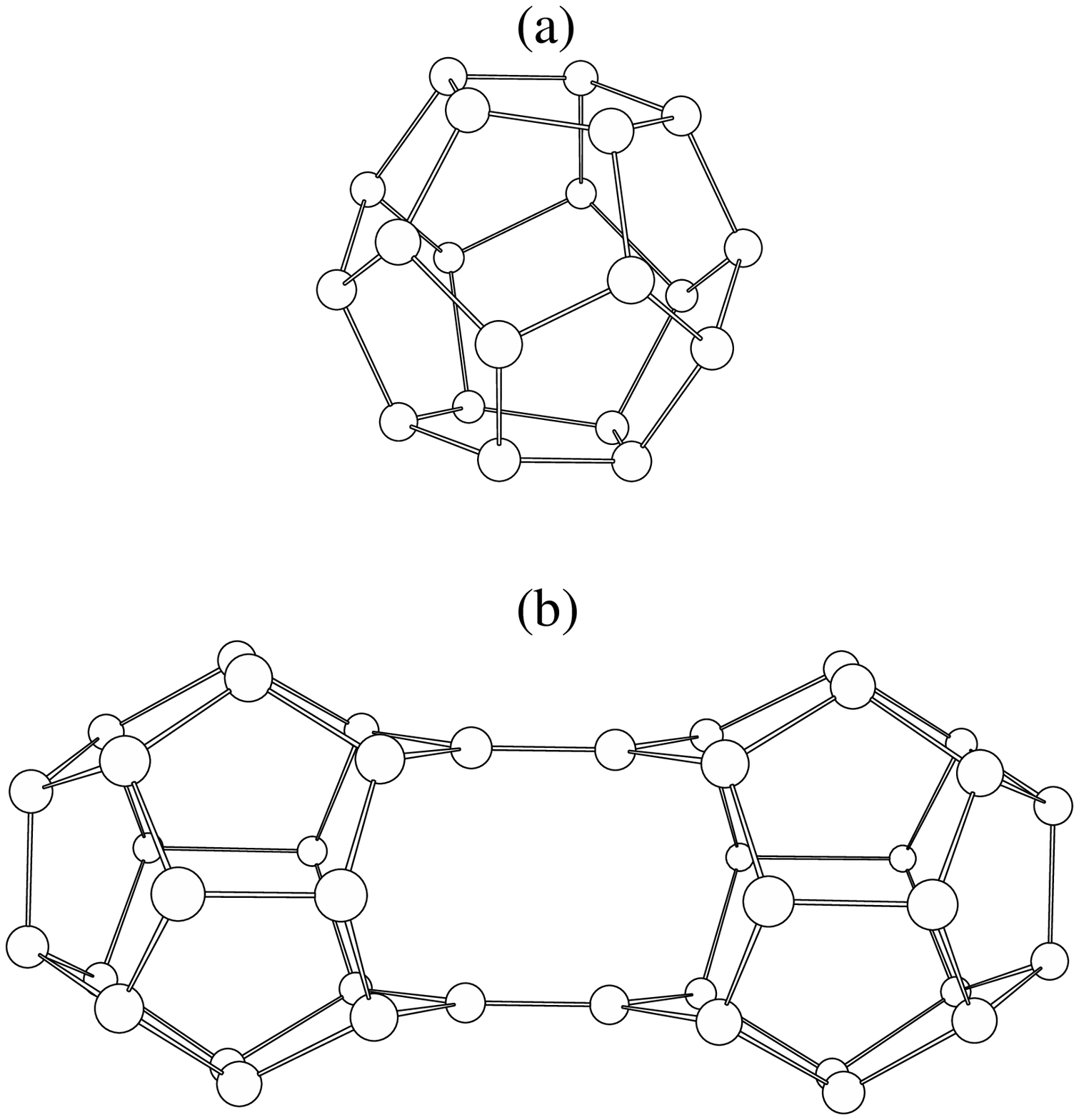}

\vskip 20mm

Fig.1. (a) Fullerene C$_{20}$; (b) {\it open}-[2+2] isomer of the
(C$_{20}$)$_2$ cluster dimer. The binding energy of fullerenes in
the dimer is $\Delta E=$ 2.470 eV/C$_{20}$.

\newpage

\includegraphics[width=15cm,height=\hsize]{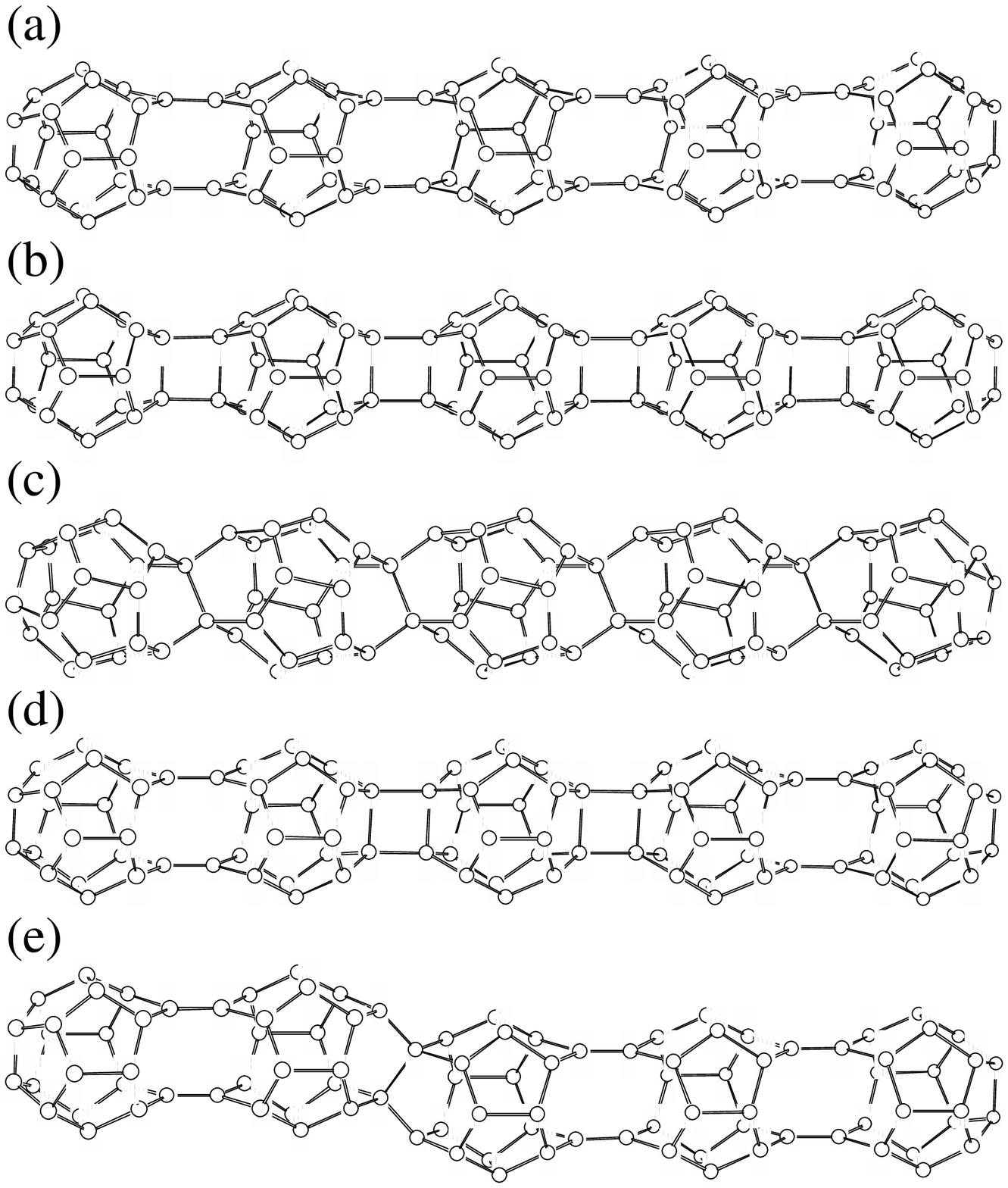}

\vskip 5mm

Fig. 2. Chains (C$_{20}$)$_5$ with the (a) {\it open}-[2+2], (b)
[2+2], and (c) twisted bonds, as well as chains with the
combination of (d) {\it open}-[2+2]  bonds and [2+2] bonds and (e)
{\it open}-[2+2] bonds and twisted bonds. The binding energies
$\Delta E$ of fullerenes in the chains are (a) 3.598, (b) 2.298,
(c) 3.511, (d) 2.973, and (e) 3.630 eV/C$_{20}$.

\newpage

\includegraphics[width=14cm,height=14cm]{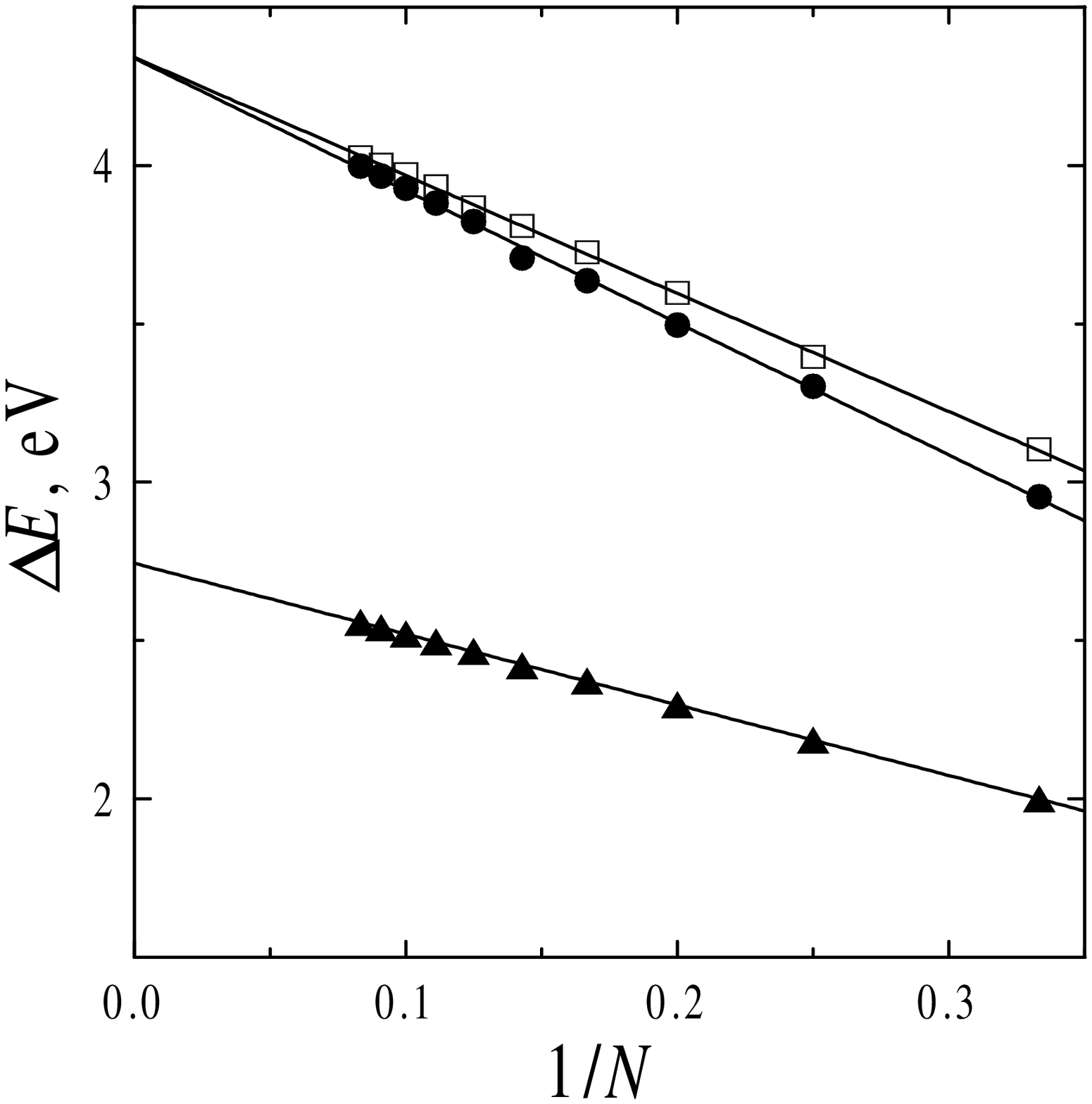}

\vskip 10mm

Fig. 3. Binding energies $\Delta E$ of C$_{20}$ fullerenes in
(Ñ$_{20}$)$_N$ chains vs. 1/$N$ for the chains with the
intercluster (triangles) [2+2] bonds, (squares) {\it open}-[2+2]
bonds, and (circles) twisted bonds, see Figs. 2a-c. The solid
lines are the rms approximation.

\newpage

\includegraphics[width=17cm,height=12cm]{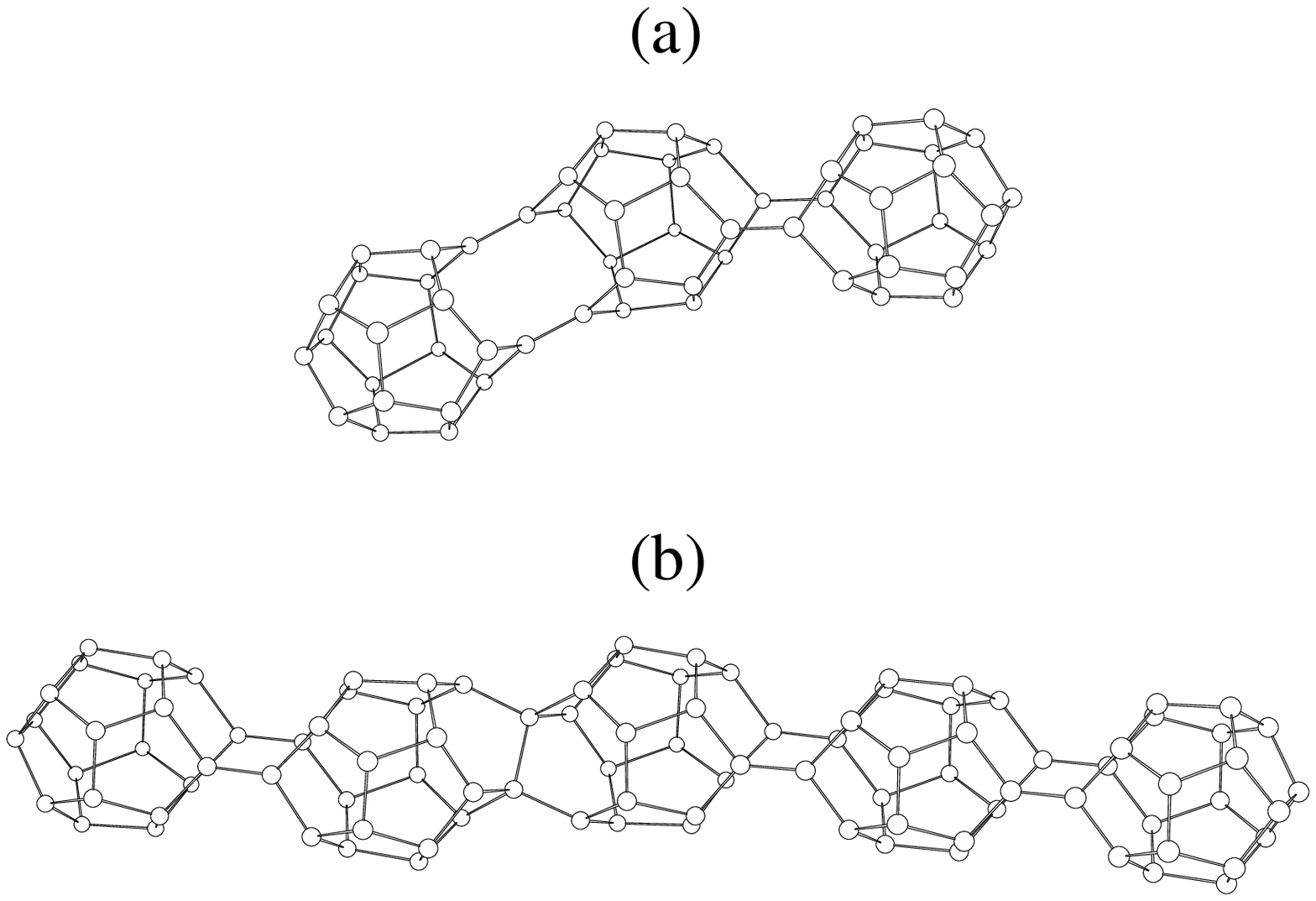}

\vskip 20mm

Fig. 4. (a) (C$_{20}$)$_3$ isomer with {\it open}-[2+2] bonds
rotated with respect to each other. (b) (C$_{20}$)$_5$ isomer
appearing due to the rotation and subsequent twisting of the {\it
open}-[2+2] bond between the second and third C$_{20}$ fullerenes.
The binding energies $\Delta E$ of fullerenes in the chains are
(a) 3.122 and (b) 3.624 eV/C$_{20}$.

\newpage

\includegraphics[width=17cm,height=18cm]{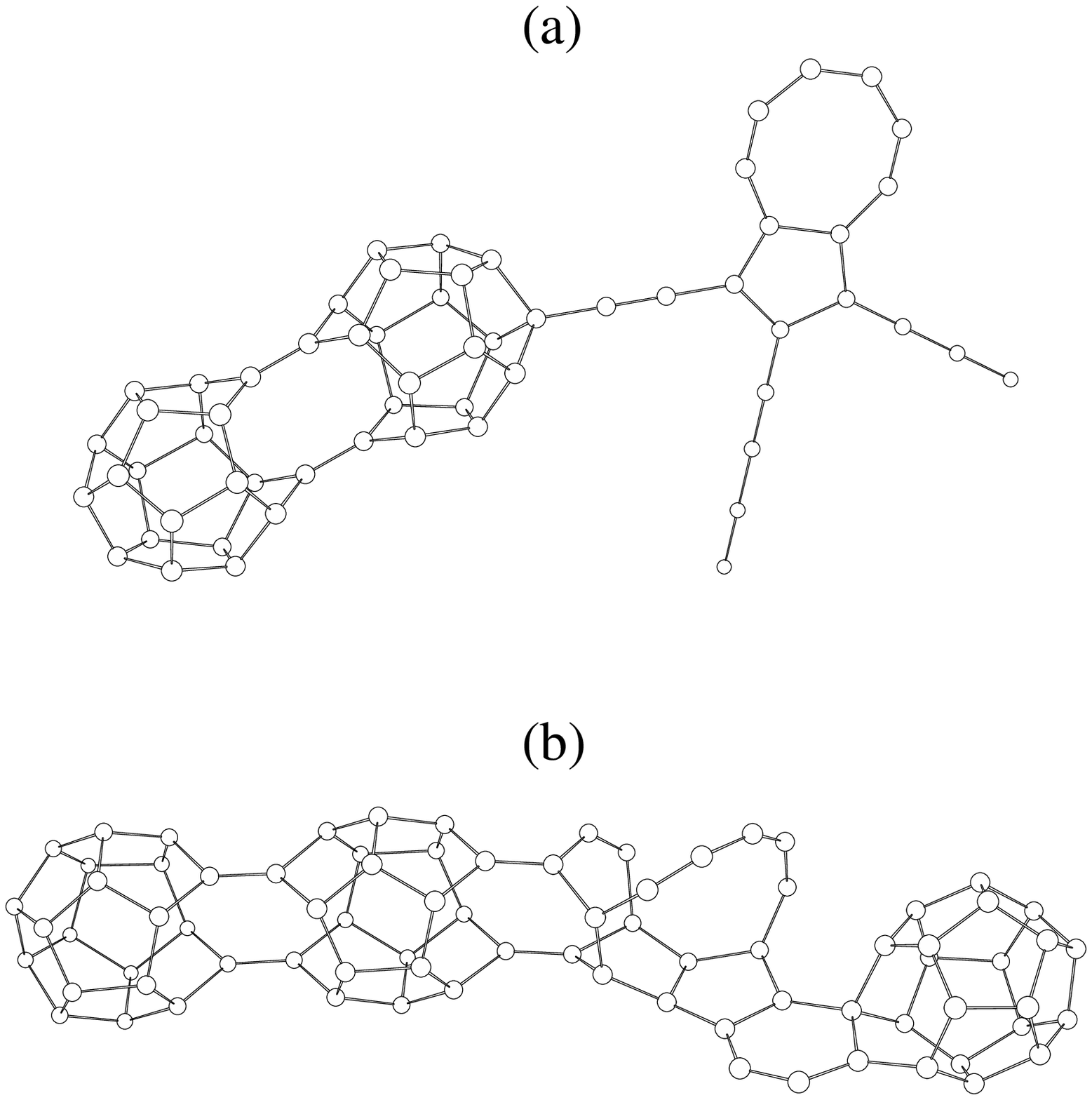}

\vskip 10mm

Fig. 5. Metastable configurations formed after the decay of one of
the C$_{20}$ fullerenes in the (a) (C$_{20}$)$_3$ and (b)
(C$_{20}$)$_4$ chains. The initial temperature $T=$ (a) 3070 and
(b) 3280 K, the lifetime before decay $\tau =$ (a) 11.6 and (b)
0.8 ps, and the binding energy after relaxation $\Delta E=$ (a)
0.790 and (b) 1.772 eV/C$_{20}$.

\newpage

\includegraphics[width=16cm,height=13cm]{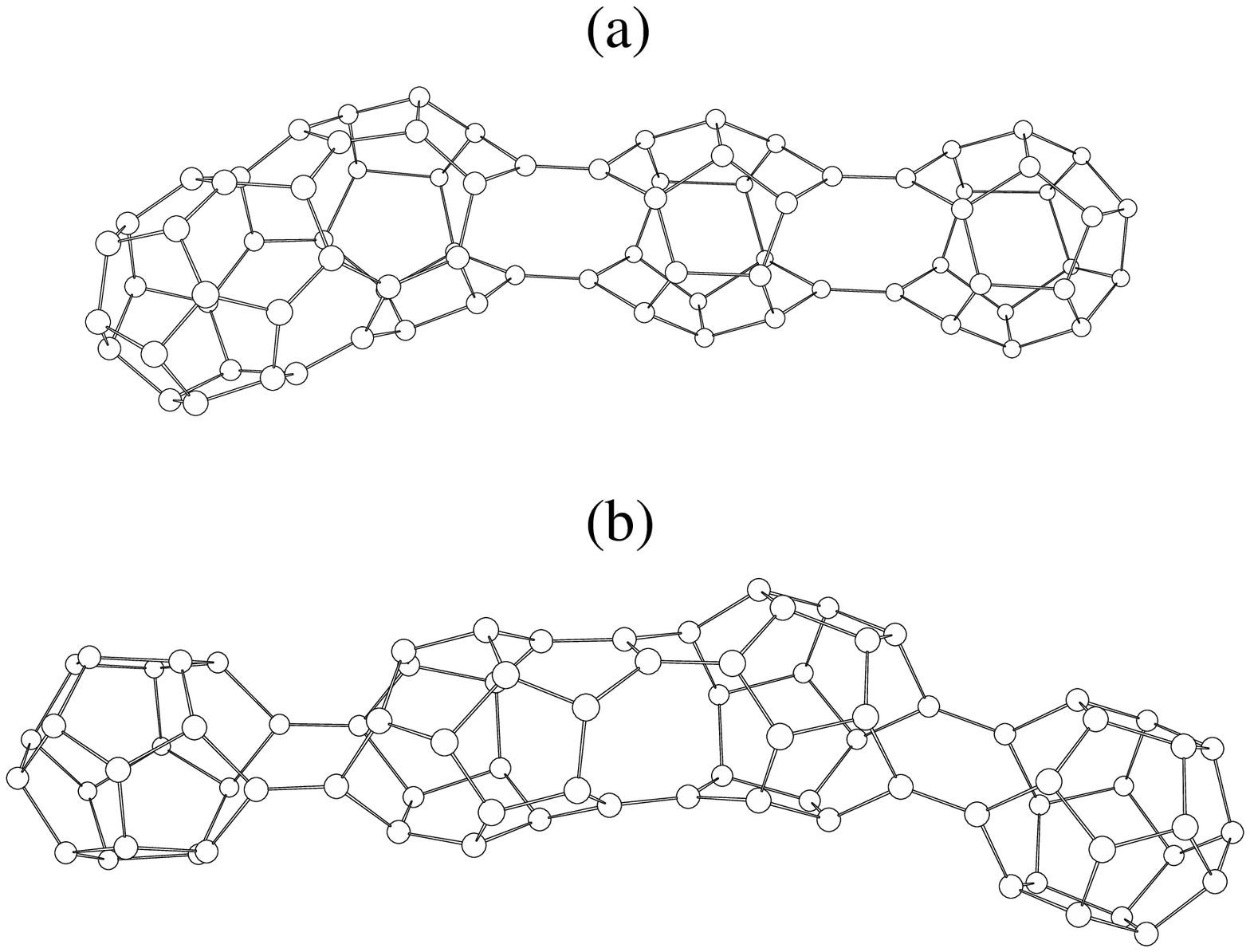}

\vskip 10mm

Fig. 6. Various configurations of the (C$_{20}$)$_4$ chain that
are formed after the coalescence of two C$_{20}$ fullerenes into a
C$_{40}$ cluster. The initial temperature $T=$ (a) 2720 and (b)
2460 K, the lifetime before coalescence $\tau =$ (a) 0.17 and (b)
26.9 ps, and the binding energy after relaxation $\Delta E=$ (a)
5.059 and (b) 3.880 eV/C$_{20}$.

\newpage

\includegraphics[width=11cm,height=19cm]{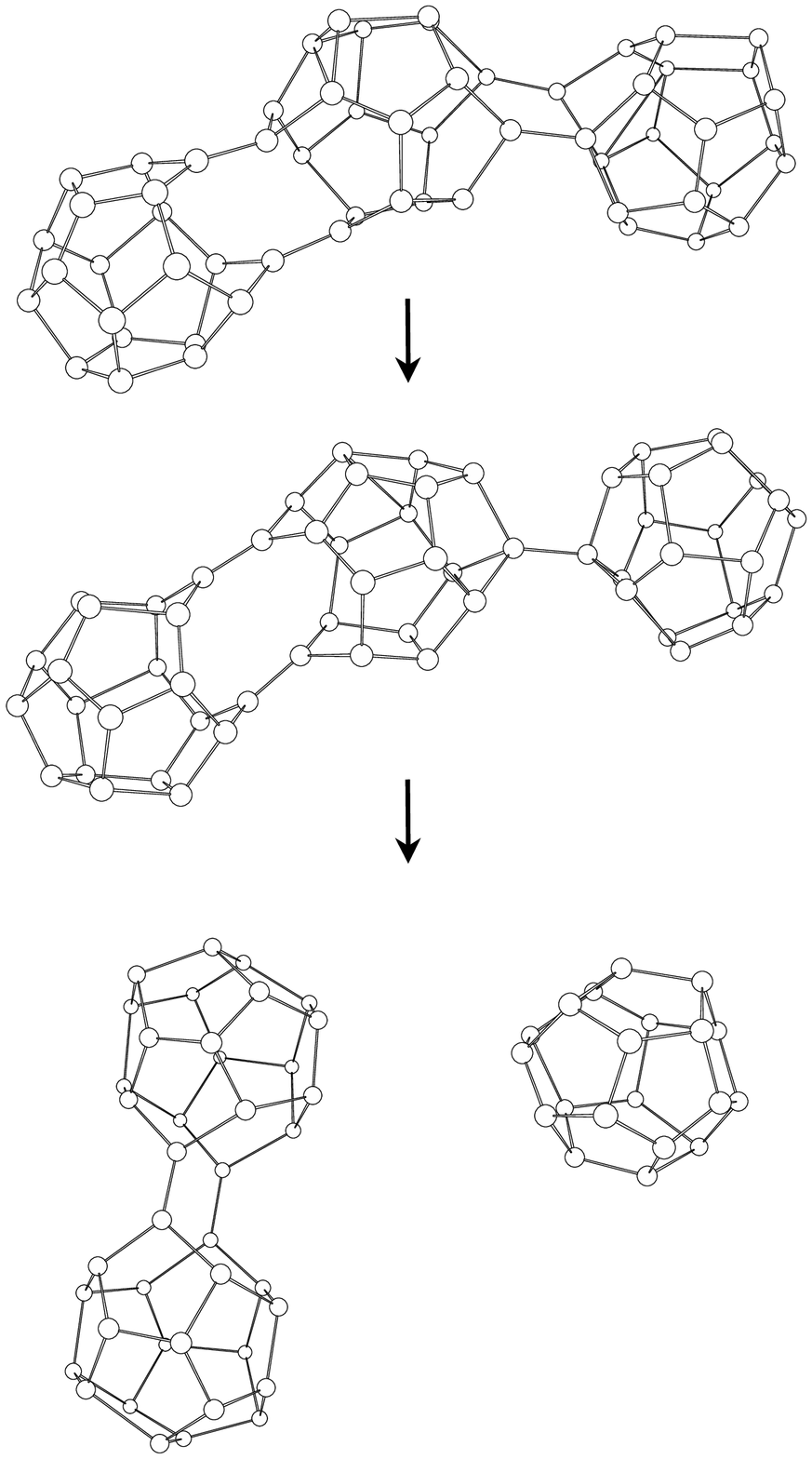}

\vskip 5mm

Fig. 7. Snapshots of the (Ñ$_{20}$)$_3$ chain (a) before the
separation of the C$_{20}$ fullerene, (b) at the intermediate
stage, and (c) after the separation. The initial temperature is
$T=2200$ K, the lifetime before the separation is $\tau =$ 1.76
ns, and the time interval between the first and third snapshots is
$\Delta t=$ 3.8 ps.

\newpage

\includegraphics[width=15cm,height=14cm]{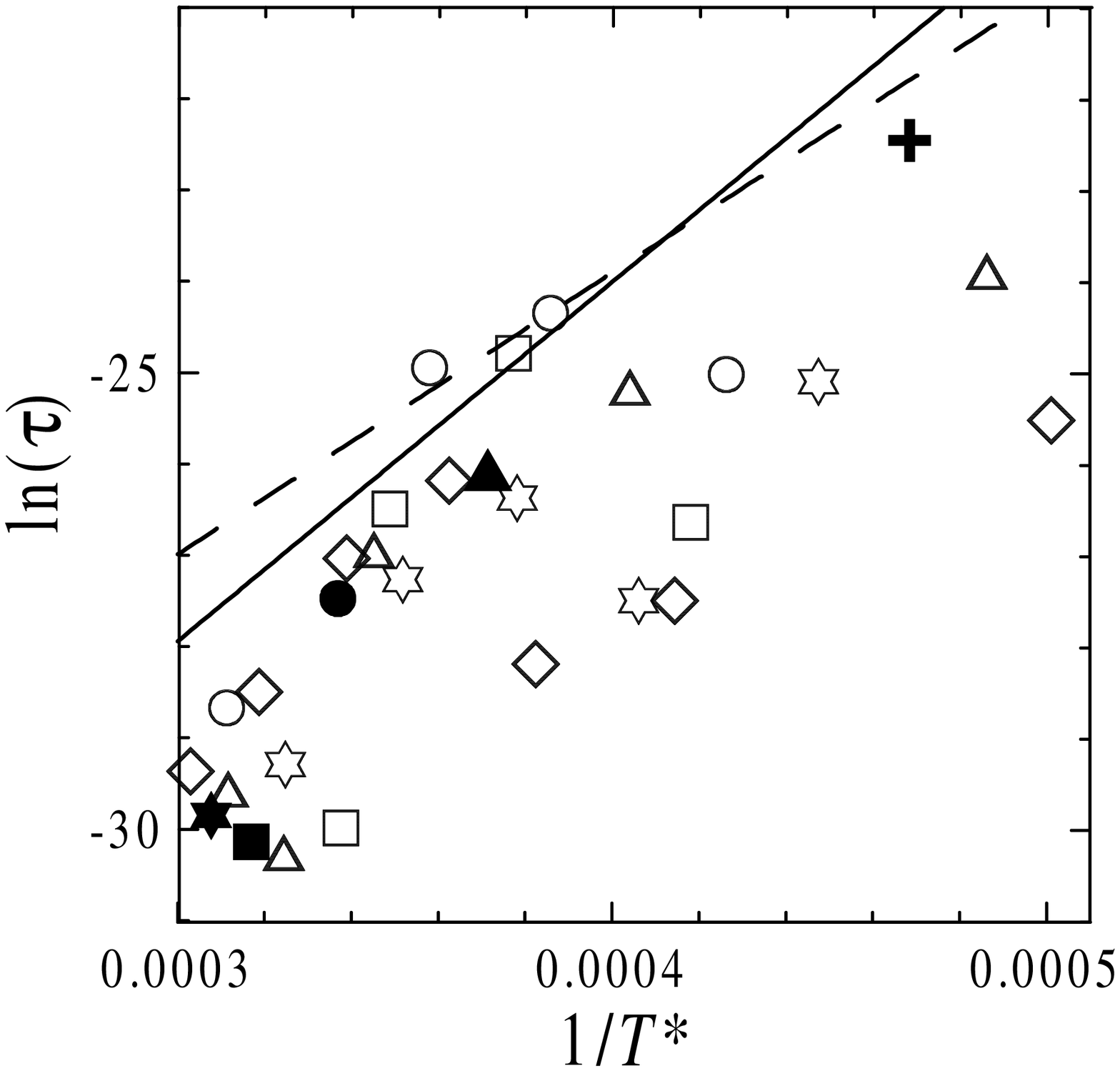}

\vskip 10mm

Fig. 8. Logarithm of the lifetime $\tau $ (second) of the
(C$_{20}$)$_N$ chains to the time of (closed points) the decay of
one of the C$_{20}$ fullerenes or (open points) coalescence of two
C$_{20}$ fullerenes into a C$_{40}$ cluster vs. the inverse
initial temperature (Kelvins) with the inclusion of the
finite-heat-bath correction, see the main text, for $N=$ (circles)
3, (squares) 4, (triangles) 5, (diamonds) 6, and (stars) 7. The
cross corresponds to the separation of one C$_{20}$ fullerene for
$N=3$. The solid and dashed lines correspond to the decay and
coalescence, respectively, of the C$_{20}$ fullerenes into the
(C$_{20}$)$_2$ dimer (from [13]).

\end{document}